\documentclass[aps]{revtex4}
\usepackage{dsfont}
\def\p  {\pi}
\def\no {\noindent}
\def\vs {\vspace}
\def\ul {\underline}
\begin{document}

\title{Higher order corrections to black hole entropy}
\author{Surhud Shrikant More}
\affiliation{Department of Physics,\\
	     Indian Institute Of Technology Bombay,\\
	     Mumbai, 400076,\\
	     India.\footnote{The work was carried out at Dept. of Physics,  University of Lethbridge, 4401 University Drive, Lethbridge, Alberta T1K 3M4, Canada}}
\email{surhudm@iitb.ac.in,surhudm@gmail.com}

\begin{abstract}
A scheme for calculating corrections to all orders to the entropy of any thermodynamic system due to statistical fluctuations around equilibrium has been developed. It is  then applied to the BTZ black hole, AdS Schwarzschild black Hole and Schwarzschild  black Hole in a cavity. The scheme that we present is a model independent scheme and hence universally applicable to all classical black holes with positive specific heat. It has been seen earlier that the microcanonical entropy of a system can be more accurately reproduced by considering a logarithmic correction to the canonical entropy function. The higher order corrections will be a step further in calculating the microcanonical entropy of a black hole.
\end{abstract}

\maketitle

\section{Introduction}
\paragraph*{}
It is well known that black holes obey laws which are analogous to the laws of thermodynamics\cite{bhal2}. The similarity of the expression
\begin{equation}
dM = T_{H} \frac{dA}{4} + \Phi dQ
\end{equation} 
with the expression in thermodynamics
\begin{equation}
dE = T dS - P dV
\end{equation} 
and the property of the nondecreasing black Hole area found by Hawking\cite{bhal1} strongly supported the Bekenstein conjecture \cite{bhal2} that the entropy of a black hole is proportional to its area. The event horizon prevents us from knowing any other information except the mass, the charge and the angular momentum of the black hole (commonly called as the ``No hair theorem"\cite{nohair}). Thus various black holes formed by different configurations of infalling matter can have the same mass, charge or angular momentum. These variables are like the thermodynamic variables of energy and pressure. In thermodynamics we have to trade information to describe the system's overall behaviour. A system can have various possible configurations leading to the same overall behaviour. This leads to the concept of entropy. It is proportional to the logarithm of the number of possible configurations.

\paragraph*{}
Various attempts have been made to explain the microscopic origin of the entropy of the black hole in terms of such degrees of freedom \cite{qg1}-\cite{qg12}. All of them use the Boltzmann definition of entropy
\begin{equation}
S = \ln \Omega(E)
\end{equation}
where $\Omega(E)$ denotes the number of different configurations a system can have and $k_{B}$, the Boltzmann's constant has been set to $1$. The degrees of freedom for a black hole are hidden from us due to the event horizon and their logarithm can be interpreted as the entropy of the black hole, in the same spirit as the entropy of a thermodynamic system.
\paragraph*{}
In statistical mechanics we have the notions of the micro-canonical and the canonical ensemble. In general one can define entropy in both these ensembles. An important thing that should be noted is the difference between the two definitions. While in the microcanonical ensemble the energy is fixed at a given point $E$, in the canonical ensemble the energy is allowed to fluctuate about a mean energy $\langle E \rangle$. If one uses the notion of entropy as information, one can conclude that the canonical entropy should be more than the microcanonical entropy. This is because in the canonical ensemble there is an additional ambiguity which remains about which states a system can take as the system can be in a state with energy close to $\langle E \rangle$ apart from the states with energy equal to $\langle E \rangle$.

\paragraph*{}
In the next section we show that the leading order difference between the canonical and the microcanonical entropy for any thermodynamic system is given by $-\frac{1}{2} \ln (CT^2)$ where $C$ is the specific heat of the system and $T$ is its temperature as has been shown earlier \cite{cft1}. This is accounted by the logarithmic corrections found out by other methods\cite{qg11} suggesting that the semiclassical arguments leading to Bekenstein Hawking entropy give us the canonical entropy. That such a correction to the canonical entropy leads closer to the microcanonical entropy also has been verified analytically and numerically in \cite{cft1}. 

\paragraph*{}
It was conjectured in (\cite{cft1} and references therein) that the next order correction should be proportional to the inverse of area\cite{bhaddas}. We develop a scheme to calculate corrections to all orders to the Bekenstein Hawking entropy using techniques of thermodynamics.

\paragraph*{}
The rest of the paper is organised as follows. In section \ref{three} we apply our results to the BTZ blackhole, the Anti de-Sitter Schwarzschild black hole and the Schwarzschild black hole in a cavity. In section \ref{four} we summarise the results obtained by us, explain the various assumptions which went in and the future directions of work.

\section{Analysis of the thermodynamic fluctuations}
\label{two}
\paragraph*{}
Consider a many body system with single particle energy spectrum $E_{n}$
completely specified by the quantum number $n$\cite{areaspec}. Thus we have 
\begin{equation}
E_{n}=f(n),~~~~~~~~n~~\epsilon~~\mathds{ Z^*}
\label{speceqn}
\end{equation}
where we assume that $f$ is a bijection from $\mathds{N}$ to $\mathds{R}$ . We also assume that $f(n)$
has a differentiable inverse $F$ such that $n=F(E)$. Let the degeneracy of the states at energy $E_{n}$ be given by $\Omega(E_{n})$. The quantum density of the system is then given by
\begin{equation}
\rho(E)=\sum_{n} \Omega(E_n) \delta (E-E_n)~.
\label{eqn0}
\end{equation}
\paragraph*{}
We use the delta function identity
\begin{equation}
\delta(E-E_n)=\delta(E-f(n))=\delta(n-F(E))|F'(E)|~,
\end{equation}
to transfer the delta function to $n$ from $E$. Here the prime denotes the first derivative of $F$ with respect to $E$. The quantum density of states is then given by 
\begin{equation}
\rho(E)=\Omega(E)|F'(E)|\sum_{n=0}^{n=\infty} \delta(n-F(E))~.
\label{eqn1}
\end{equation}
\paragraph*{}
The Poisson summation formula 
\begin{equation}
\sum_{n=-\infty}^{\infty} \delta(n-F(E))=\sum_{k=-\infty}^{\infty}\exp(2i\pi kF(E))~.
\end{equation}
leads to  
\begin{equation}
\sum_{n=0}^{\infty} \delta(n-F(E))=1+2\sum_{k=1}^{\infty} {\rm cos}(2\pi kF(E))~.
\label{eqn2}
\end{equation}
considering the fact that $F(E) \geq 0$. Using Eq.(\ref{eqn1}) and the Poisson summation formula we get
\begin{equation}
{\rho}(E)=\Omega(E)|F'(E)|\left( 1+2\sum_{k=1}^{\infty} {\rm cos}(2\pi kF(E)) \right)
\end{equation} 
\paragraph*{}
Note that the notation and the derivation till here follows the scheme of \cite{dasbhaduritran}. We thus obtain 
\begin{equation}
\tilde{\rho}(E)=\Omega(E)|F'(E)|~,
\label{eqn3}
\end{equation}
where $\tilde{\rho}(E)$ denotes the averaged smoothed density of states. Now we consider the case of $N$ noninteracting particles constituting the many body system. The microcanonical entropy is just then the natural logarithm of the degeneracy $\Omega(E_{n})$ which is
\begin{equation}
S = \ln [\tilde{\rho}(E)|F'(E)|^{-1}]~,
\label{eqn4}
\end{equation} 
\paragraph*{}
If one assumes that the function $\rho(E)$ is smooth and neglects the oscillating components which arise from the discreteness of the energy levels (which can be done for a macroscopic system with large energy $E$), then
\begin{equation}
\tilde{\rho}(E) = \rho(E)
\end{equation}
We now consider the canonical ensemble to calculate the density of states. The canonical $N$ particle partition function is given by 
\begin{equation}
Z_{c}(\beta) = \sum_{n}\Omega(E_n) \exp(-\beta E_n)~= 
\int_0^{\infty} \rho(E) \exp(-\beta E) dE~, 
\label{eqn5}
\end{equation} 
where $\rho(E)$ is defined by Eq.(\ref{eqn0}). The canonical ensemble is used as a tool here to extract the canonical density of states\cite{york}. We observe that the canonical partition function is the Laplace transform of the density of states. So we obtain $\rho(E)$ by Laplace inversion with respect to $\beta$\cite{lap1,lap2}, which will be our variable along the imaginary axis. Thus 
\begin{equation}
\rho(E) = \frac{1}{2\pi i} \int_{-i \infty}^{i \infty} Z_{c}(\beta) e^{\beta E} d\beta
\end{equation} 
which can be rewritten as
\begin{equation}
\rho(E) = \frac{1}{2\pi i} \int_{-i \infty}^{i \infty}  e^{\ln(Z_{c}(\beta))+\beta E} d\beta
\label{eqn100}
\end{equation} 
\paragraph*{}
Here $E$ is the canonical ensemble energy average $E$. We expect that the function $S(\beta) = \ln(Z_{c}(\beta))+\beta E $ has an extremum which the thermodynamic system will tend to. Let the extremum be at $\beta=\beta_{0}$. One can expand the function $S(\beta)$ in a Taylor series around $\beta_{0}$ as
\begin{equation}
S(\beta)=\ln(Z_{c}(\beta))+\beta E=S_0(E)+\frac{1}{2!}\left(\frac{\partial^2}{\partial\beta^2}\ln Z_{c}\right)_{\beta_{0}}(\beta-\beta_{0})^2+\frac{1}{3!}\left(\frac{\partial^3}{\partial\beta^3}\ln Z_{c}\right)_{\beta_{0}}(\beta-\beta_{0})^3+....
\label{eqn101}
\end{equation}
Since the first derivative of $S$ with respect to $\beta$ vanishes at $\beta_{0}$  using Eq.(\ref{eqn100}) and Eq.(\ref{eqn101}), we get
\begin{equation}
\rho(E)=\frac{1}{2\pi i} \int_{-i \infty}^{i \infty} e^{S_0(E)}e^{\frac{1}{2!}\alpha_{2}^{2} (\beta-\beta_{0})^2 + \frac{1}{3!} \alpha_{3} (\beta-\beta_{0})^3 + \frac{1}{4!} \alpha_{4} (\beta-\beta_{0})^4 + ... }
\end{equation} 
Note that here we have assumed that the second derivative is positive and defined as :
\begin{equation}
\alpha_{2}^{2}=\left(\frac{\partial^2}{\partial\beta^2}\ln Z_{c}\right)_{\beta_{0}}
\end{equation} 
while for the rest $n$, $\alpha_{n}$ denotes the $n$-$th$ derivative of $\ln Z_{c}$.
\begin{equation}
\alpha_{n}=\left(\frac{\partial^n}{\partial\beta^n}\ln Z_{c}\right)_{\beta_{0}}
\end{equation}
\paragraph*{}
Changing variable from $(\beta-\beta_{0})$ to $iy$, we get
\begin{equation}
\rho(E)= \frac{1}{2\pi} e^{S_0(E)} \int_{-\infty}^{\infty} e^{-\frac{1}{2}\alpha_{2}^{2}y^2} e^{ \sum_{n=3}^{\infty} \frac{\alpha_n (i)^n y^n}{n!} }
\end{equation} 
The second exponential can be expanded as a series noting that for a large thermodynamic system the exponent is small to give
\begin{equation}
\rho(E)= \frac{1}{2\pi} e^{S_0(E)} \int_{-\infty}^{\infty} e^{-\frac{1}{2}\alpha_{2}^{2}y^2}\left( \sum_{m=0}^{\infty} \frac{1}{m!}\left(\sum_{n=3}^{\infty} \frac{\alpha_n (i)^n y^n}{n!}\right)^m \right)
\label{meqn}
\end{equation}  
\paragraph*{}
As we know that the L.H.S is a real quantity we can safely ignore the terms giving imaginary quantities in the R.H.S. The complete expansion will involve the multinomial expansion. We use terms only up to second order in $m$ here \textit{i.e} up to $m=2$, noting that the highest order in the neglected next terms is $O(y^{12})$ on account of considering only non-imaginary terms.(The cube of the term with $n=3$ does not contribute being imaginary.) To simplify matters just consider the expansion which we did in the last step.
\begin{equation}
e^{ \sum_{n=3}^{\infty} \frac{\alpha_n (i)^n y^n}{n!} }= \left( \sum_{m=0}^{\infty} \frac{1}{m!}\left(\sum_{n=3}^{\infty} \frac{\alpha_n (i)^n y^n}{n!}\right)^m \right)
\end{equation}
The terms which contribute real terms are then (up to $m=2$)
\begin{equation}
1 + \sum_{n=2}^{\infty} \frac{ \alpha_{2n}(-1)^n y^{2n} }{ (2n)! } + \frac{1}{2!} \left\{ \sum_{n=3}^{\infty} \sum_{m=3}^{\infty} \frac{ \alpha_{n} \alpha_{m}(i)^{n+m} y^{n+m} }{n! m! } \right\} ~~~~~~n+m=2k~~~~ k~~\epsilon~~\mathds{N}
\label{appreqn}
\end{equation}
\paragraph*{}
Using  Eq.(\ref{meqn}) and the standard integrals 
\begin{equation}
\int_{-\infty}^{\infty} e^{ -\frac{1}{2}\alpha_{2}^{2}y^2 } dy = \sqrt{\frac{2\pi}{\alpha_{2}^{2} }}
\end{equation}
and
\begin{equation}
\int_{-\infty}^{\infty} e^{ -\frac{1}{2}\alpha_{2}^{2}y^2 } y^{2n} dy = \sqrt{\frac{2\pi}{\alpha_{2}^{2} }} \frac{(2n-1)!!}{\alpha_{2}^{2n} }
\end{equation}
we get the integrated expression for $\rho(E)$ as follows
\begin{equation}
\rho(E)= \frac{1}{2\pi} e^{S_0(E)} \sqrt {\frac{2\pi}{\alpha_{2}^{2}} }\left[ 1 + \sum_{n=2}^{\infty} \frac{\alpha_{2n}(-1)^n}{(2n)!! \alpha_{2}^{2n}} + \frac{1}{2!}  \sum_{n=3}^{\infty} \sum_{m=3}^{\infty} \frac{\alpha_{n}\alpha_{m}(-1)^k (2k-1)!! }{ n! m! \alpha_{2}^{2k} } + ... \right]
\end{equation}
The above derivation is in the spirit of corrections to the saddle point formula\cite{bhaduri,hoare}. Here we would like to contrast our approach with the approach in \cite{muin} where the $\beta^{3}$ term has been ignored as it contributes an imaginary term. Hence some extra contributions are missing following the approach in \cite{muin}.
\paragraph*{}
In the limit of a macroscopic system with a very large value of energy, we neglect the oscillating components in $\rho (E)$ and put it equal to $\tilde{\rho} (E)$. To get the microcanonical canonical entropy we take the logarithm of both sides to get
\begin{equation}
\ln(\rho(E))=\ln(\Omega(E))+\ln(F(E))=S(E)+\ln(F(E))
\end{equation} 
\begin{equation}
S(E)=S_{0}(E) - \ln(F(E)) - \frac{1}{2}\ln(\alpha_{2}^{2}) + \ln \left( 1 + \sum_{n=2}^{\infty} \frac{\alpha_{2n}(-1)^n}{(2n)!! \alpha_{2}^{2n}} + \frac{1}{2!}  \sum_{n=3}^{\infty} \sum_{m=3}^{\infty} \frac{\alpha_{n}\alpha_{m}(-1)^k (2k-1)!! }{ n! m! \alpha_{2}^{2k} } + O(\frac{\alpha_{4}^{3}}{\alpha_{2}^{12}})\right)
\end{equation}
Next we use the approximation
\begin{equation}
\ln (1+x) \sim x
\end{equation} 
to get the final expression 
\begin{equation}
S(E)=S_{0}(E) - \ln(F(E)) - \frac{1}{2}\ln(\alpha_{2}^{2}) + \sum_{n=2}^{\infty} \frac{\alpha_{2n}(-1)^n}{(2n)!! \alpha_{2}^{2n}} + \frac{1}{2!}  \sum_{n=3}^{\infty} \sum_{m=3}^{\infty} \frac{\alpha_{n}\alpha_{m}(-1)^k (2k-1)!! }{ n! m! \alpha_{2}^{2k} } + O(\frac{\alpha_{4}^{3}}{\alpha_{2}^{12}})
\label{finent}
\end{equation} 
This is our formula to calculate the corrections to black hole entropy which is valid for any thermodynamic system. Thus in addition to the logarithmic correction the microcanonical entropy formula has, a logarithmic correction dependent on the energy spectrum of the black hole. Such corrections and their possible origins have been studied in \cite{dasbhaduritran}.
%\newpage{}
\section{Applications to Black hole entropy}
\label{three}

We now apply Eq.(\ref{finent}) to find out entropy corrections to the BTZ blackhole, the Anti de-Sitter schwarzschild black hole and the Schwarzschild black hole in a cavity.
%\no
\subsection{BTZ Black Hole}
\paragraph*{}
Consider a non-rotating BTZ black hole in three dimensions with metric \cite{btz}
\begin{equation}
ds^2 = - \left( \frac{r^2}{\ell^2} - 8 M \right) dt^2
+ \left( \frac{r^2}{\ell^2} - 8 M \right)^{-1}  dr^2
+ r^2 d\theta^2 ~~.
\end{equation}
\paragraph*{}
The black hole behaves as a thermodynamic system with the thermodynamic quantities namely its Hawking temperature and Bekenstein-Hawking entropy given by the expressions
\begin{eqnarray}
T_H = \frac{r_+}{2\pi \ell^2} = \left[ \frac{1}{\p^2\ell^2}\right]S_0\\
S_0 = \frac{2\pi r_+ }{4}
\end{eqnarray}
where the horizon radius is denoted by $r_{+}$ and given by $r_{+}= \sqrt{8G_3M}\ell$, $M$ being the mass of the black hole and $\ell$ is related to the cosmological constant by $\Lambda=-\frac{1}{\ell^2}$. It can be shown that $M$, $T_{H}$ and $S$ satisfy the first law of Thermodynamics.
\paragraph*{}
Hence we have
\begin{equation}
S(\beta)= \pi^{2}\ell^{2} \frac{1}{\beta_{H}} = \frac{K}{\beta_{H}} 
\end{equation}
where K is a constant.
\paragraph*{}
Using this expression we can calculate the required $\alpha_{n}$'s.
\begin{eqnarray}
\alpha_{2}^{2} = 2 \frac{K}{\beta_{H}^{3}} = 2 \frac{S_{0}^3}{K^2}
\label{surhud1}\\
(\alpha_{n})_{n \geq 3} = (-1)^n\frac{K n!}{\beta_{H}^{n+1}} = (-1)^{n}\frac{n!S_{0}^{n+1}}{K^{n}}
\label{surhud2}
\end{eqnarray}
Thus using Eq.(\ref{finent}) we find that in addition to the logarithmic correction arising due to the energy spectrum we have another logarithmic correction given by 
\begin{equation}
- \frac{1}{2}\ln(\alpha_{2}^{2}) = -\frac{3}{2} \ln S_{0} + K_{1}
\end{equation}
where $K_{1}$ is a constant. This reproduces the correction in \cite{cft1}. 
\paragraph*{}
Now we proceed towards the higher order terms :
\begin{equation}
Higher~~order~~corrections = \sum_{n=2}^{\infty} \frac{\alpha_{2n}(-1)^n}{(2n)!! \alpha_{2}^{2n}} + \frac{1}{2!}  \sum_{n=3}^{\infty} \sum_{m=3}^{\infty} \frac{\alpha_{n}\alpha_{m}(-1)^k (2k-1)!! }{ n! m! \alpha_{2}^{2k} } 
 \end{equation}
We consider these terms one by one and use Eq.(\ref{surhud1}) and Eq.(\ref{surhud2}) to get
\begin{eqnarray}
\sum_{n=2}^{\infty} \frac{\alpha_{2n}(-1)^n}{(2n)!! \alpha_{2}^{2n}} = \sum_{n=2}^{\infty} \left(\frac{-1}{2}\right)^n \frac{(2n-1)!!}{S_{0}^{n-1}}\\
\sum_{m=3}^{\infty} \frac{\alpha_{n}\alpha_{m}(-1)^k (2k-1)!! }{ n! m! \alpha_{2}^{2k} } = \sum_{n=3}^{\infty} \sum_{m=3}^{\infty} \frac{(-1)^k}{2^{k+1}}\frac{(2k-1)!!}{S_{0}^{k-2}}
\end{eqnarray}
Thus we have corrections as polynomials in $\frac{1}{S_{0}}$\cite{log1}-\cite{log6}. The leading term in them being $\frac{1}{S_{0}}$ appearing with the coefficient occurring for n=2 for the first of the equations above and $k=3$ translating to $m=3$, $n=3$ for the following equation. Thus we get
\begin{equation}
S(E)= S_{0}(E) -  \frac{3}{2}\ln(S_{0}) - \frac{3}{16} \frac{1}{S_{0}} + ... -\ln(F(E)^{-1})
\label{entbtz}
\end{equation} 
\paragraph*{}
Also note that we can trust the factor of the $S_{0}^{-1}$ term as the $O(y^{12})$ terms neglected in the expansion of the exponential in Eq.(\ref{appreqn}) give a term $\frac{1}{S_{0}^3}$

\subsection{Anti de-Sitter Schwarzschild Black Hole}
\paragraph*{}
Consider a d-dimensional Anti de-Sitter Schwarzschild black hole.We parameterise the cosmological constant in $d$-dimensions in terms of the length scale following \cite{cft1} as 
\begin{equation}
\Lambda=-\frac{(d-1)(d-2)}{2\ell^{2}}
\end{equation}
The AdS Schwarzschild black hole has the metric 
\begin{equation}
ds^2 = - \left(1-\frac{16\pi M}{(d-2)\Omega_{d-2}r^{d-3}} + \frac{r^2}{\ell^2} \right) dt^{2} + \left(1-\frac{16\pi}{(d-2)\Omega_{d-2}r^{d-3}} + \frac{r^2}{\ell^2} \right)^{-1} dr^{2} + r^{2}d\Omega_{d-2}^{2}
\end{equation} 
The parameter $M$ is the conserved charge associated with time-like Killing vector. In terms of the horizon radius then the entropy and the Hawking temperature are given by
\begin{equation}
S_0 = \frac{\Omega_{d-2}r_{+}^{d-2} }{4 } \label{eqn1sch}
\end{equation}
\begin{equation}
T_{H} = \frac{ (d-1)r_{+}^{2} + (d-3)\ell^{2} }{4\pi \ell^{2} r_{+} } 
= \left[ \frac{(d-3)}{4\pi} \frac{\Omega_{d-2}}{4} \right] \left[ 1 + \frac{(d-1)r_{+}^{2}}{(d-3)\ell^{2}}\right] S_{0}^{-\frac{1}{(d-2)}} \equiv k_{1}^{-1} S_{0}^{\frac{1}{(d-2)}}
\end{equation}
where $k_{1}$ is a constant. The last approximation arises considering the large $r_{+}$ limit so that $ r_{+} \gg \ell $ which is the so-called high temperature limit.
 The specific heat then turns out to be \cite{cft1} positive for a wisely chosen value of $\ell$. In the limit of $\ell \rightarrow 0 $ we get the specific heat to be positive for $d \geq 3$ and given by the expression
\begin{equation}
C = (d-2) S_{0}
\end{equation}
\paragraph*{}
The relation between $S_{0}$ and $\beta_{H}$ is then given by
\begin{equation}
S_{0} = k_{1}^{d-2} \beta_{H}^{-(d-2)} = k_{2} \beta_{H}^{-(d-2)}
\end{equation} 
where $k_{2}$ is a constant.
Using this we calculate the corrections to the entropy. For the first order corrections we need the second derivative of the entropy with respect to inverse temperature. In our notation 
\begin{eqnarray}
\alpha_{2}^{2} = k_{2}(d-2)(d-1)\beta_{H}^{-d} \\
\alpha_{3} =  k_{2}(d-2)(d-1)(-d) \beta_{H}^{-(d+1)}\\
\alpha_{4} =  k_{2}(d-2)(d-1)(d)(d+1) \beta_{H}^{-(d+2)}
\end{eqnarray}
And with the usual notations that has been followed up till now we have,
\begin{eqnarray}
\frac{\alpha_{2n}}{\alpha_{2}^{2n}}\propto S_{0}^{1-n}\\%verifiedtobecorrect.09:30PM26/5/04
\frac{\alpha_{n}\alpha_{m}}{\alpha_{2}^{2k}} \propto S_{0}^{2-k} 
\end{eqnarray} 
the neglected term $O(y^{12})$ in the exponential expansion goes here as $O(\frac{1}{S_{0}^{3}})$ again as can be verified and thus is less than the one we are talking about. Thus the corrected entropy up to order $\frac{1}{S_{0}}$ is 
\begin{equation}
S = S_{0} - \ln(F(E)^{-1}) - \frac{(d)}{2(d-2)} \ln S_{0} +\frac{d(3-2d)}{24(d-2)(d-1)}\frac{1}{S_{0}}
\label{entads}
\end{equation} 
For example for a four dimensional Anti de-Sitter Black hole (d=4) we find that
\begin{equation}
S = S_{0} - \ln(F(E)^{-1}) - \ln S_{0} +\frac{-5}{36}\frac{1}{S_{0}}
\label{entadsspl}
\end{equation} 

\subsection{Schwarzschild in a cavity}
\paragraph*{}
A Schwarzschild black hole has a negative specific heat but a thermodynamically stable Schwarzschild black hole can be obtained within a finite cavity by immersing it in a thermodynamic bath\cite{cage}. Let us find what are the higher order corrections to the entropy of such a black hole. 

\paragraph*{}
Again we proceed the same way as the cases above. For such a black hole we have 
\begin{eqnarray}
S=4\pi M^{2}\\
\beta=8\pi M \sqrt{\left(1-2\frac{M}{R}\right)}
\end{eqnarray} 
where R is the radius of the spherical cavity in which the black hole is enclosed.
\paragraph*{}
Now note that
\begin{equation}
\frac{\partial^{2}}{\partial \beta^{2}}S = \frac{\partial}{\partial \beta}\left(\frac{\partial}{\partial \beta}S\right)=-\frac{\partial}{\partial \beta}\langle E \rangle = T^{2}\frac{\partial}{\partial T}\langle E \rangle = T^{3}\frac{\partial}{\partial T}S
\end{equation}

\paragraph*{}
One more relation to be noted is that derivable from the expression for $\beta$,
\begin{equation}
\frac{\partial}{\partial \beta}M= \frac{\sqrt{R}}{8\pi}\frac{\sqrt{(R-2M)}}{(R-3M)}
\end{equation} 

\paragraph*{}
Using these equations we get
\begin{eqnarray}
\alpha_{2}^{2}=\frac{1}{8\pi}\frac{R}{3M-R}\\
\alpha_{3}=-\frac{3}{64\pi^{2}}\frac{R^{\frac{3}{2}} \sqrt{R-2M} }{(R-3M)^{3}}\\
\alpha_{4}=\frac{-3}{(8\pi)^{3}}\frac{R^{2}(9M-4R)}{(R-3M)^{3}}
\end{eqnarray}

Plugging in Eq.(\ref{finent}). The final Expression comes out to be 
\begin{equation}
S = S_{0} -\ln(F(E)^{-1})  + \frac{1}{2}\ln(4\pi) + \frac{1}{2}\ln\left(\frac{3}{R} \sqrt{\frac{S_{0}}{\pi}} - 2 \right) - \frac{7}{16}\frac{(8R\sqrt{\pi}-9\sqrt{S_{0}})}{(3\sqrt{S_{0}}-2R\sqrt{\pi})^{3}}
\end{equation} 

This expression till first order was derived in \cite{akbardas}. One should note here that the first order corrections are not of the type $ln(S_0)$. The higher order corrections are also not of the form $1/S_0$ but they have the same order as $1/S_0$.
 
\section{Discussions}
\label{four}
In this paper we have developed a scheme to calculate the corrections to any order for a thermodynamic system. The higher order corrections arise when one expands the canonical entropy function around the extremum. The correction when applied to the BTZ black hole in the AdS Schwarzschild black hole results in terms that can be arranged to be a polynomial in $1/S_{0}$. For the Schwarzschild black hole in a cavity the correction higher in order than the logarithmic correction is of order $1/S_0$. The scheme used is universally applicable as it is model independent. 
%Here one can raise a doubt whether the quantum fluctuations around the equilibrium would dominate the thermodynamic fluctuations. But one should note that where quantum corrections have been calculated they are comparable, order by order, to the calculations done in this paper\cite{log1}.

\paragraph*{}
Since no quantum ideas were used anywhere in the derivation, the corrections calculated are corrections to the microcanonical entropy in a classical theory. Where such corrections have been found using quantum gravity theories, the leading order corrections are logarithmic and hence greater than the higher order corrections ($1/S$) found here~\cite{log1}.

Now we would like to bring into light some assumptions that were made while doing the calculations.

\paragraph*{}
 The Bekenstein Hawking entropy is taken to be the canonical entropy and it is used while calculating the derivatives for getting the corrections. We expect the canonical entropy to have an extremum to which the system will settle to. But this is not so for the Bekenstein Hawking Entropy. We demonstrate this by a specific example. The Bekenstein Hawking entropy for the BTZ black hole has a $\frac{1}{\beta}$ dependence and hence the extremum corresponds to $\beta=\infty$. To the first approximation the canonical entropy and the Bekenstein Hawking Entropy agree \cite{banados}. The way out of this situation might be as pointed out in \cite{cft1} inspired by Conformal Field Theory that the complete entropy function may be having other terms which go to zero in the limit that the black hole has mass far greater than the Planck Mass; but the function then has a saddle point due to those extra terms. One such proposed function was

\begin{equation}
S(\beta) = a\beta^m + \frac{b}{\beta^{n}}
\end{equation} 
\paragraph*{}
In such a case the extremum occurs at
\begin{equation}
\beta_{0}=\left( \frac{nb}{ma}\right)^{\frac{1}{m+n}}
\end{equation} 
If we proceed along the lines of \cite{cft1} and expand $S$ around $\beta_{0}$ we get 
\begin{equation}
S(\beta) = \alpha(a^{n}b^{m})^{\frac{1}{(m+n)}} + \frac{1}{2} \gamma (a^{n+2} b^{m-2})^{\frac{1}{m+n}} (\beta- \beta_{0})^{2}  + \delta (a^{n+3}b^{m-3})^{\frac{1}{(m+n)}} (\beta- \beta_{0})^{3} + \zeta (a^{n+4}b^{m-4})^{\frac{1}{(m+n)}} (\beta- \beta_{0})^{4} + ...
\end{equation} 
where $\alpha, \gamma, \delta, \zeta$ are functions of $m$ and $n$.
\paragraph*{}
We then proceed to use Eq.(\ref{finent}) to get the entropy
\begin{equation}
S = S_{0} -\frac{1}{2} \ln(S_{0}T_{H}^{2}) + f(m,n)\frac{1}{S_{0}}  + ...
\end{equation} 
thus verifying the prediction in \cite{cft1} that the subleading order terms have dependence on $m$ and $n$ (order unity) and also that such an entropy function has a subleading order correction of the form $\frac{1}{S_{0}}$. One should note that the  logarithmic corection also has a term involving $m$ and $n$ and it has been separated out.
 
 The analysis shall go through only under the circumstances of having such an entropy function to provide for an extremum and it shall breakdown when the Planck mass limit is approached.

\paragraph*{} 
 The logarithmic correction due to a non-uniform energy spectrum of the black hole results in a logarithmic correction in addition to the other corrections. Such a correction was first noted in \cite{gourmed}. It would be wrong to discard such spectral factors apriori without having a complete theory of spacetime fluctuations. We have left this factor in the final expression so as to allow for such conditions if they need to be considered. This factor can prove important in deciding whether the system is thermally stable or not as has been shown in \cite{majumchat}.
 
\paragraph*{}
Because the BTZ entropy function can be obtained from AdS Schwarzschild one by putting $d=3$ one can verify that the $\frac{1}{S_{0}}$ correction term in Eq.(\ref{entads}) is the same as in Eq.(\ref{entbtz}) if one substitutes $d=3$. It is interesting to note that for the BTZ blackhole, the $\frac{1}{S_{0}}$ correction that we found out is negative. The $\frac{1}{S_{0}}$ correction is also negative for the AdS Schwarzschild black hole for $d \geq 3$. At this moment it would be bold to conjecture the universality of this sign though. 
 
\paragraph*{}
Also to be noted is the fact that the order of the neglected terms in Eq.(\ref{finent}) is dimensionally of the order $\frac{1}{S_{0}^{3}}$. This means that corrections to the order of $\frac{1}{S_{0}^{2}}$ can also be found out using Eq.(\ref{finent}). Similarly all the higher order corrections can be calculated using Eq(\ref{meqn}).

\paragraph*{}
We have seen that the thermal fluctuations result in corrections to the black hole entropy. As a result the Cardy Verlinde formula receives corrections via the holographic relations. The effect of the logarithmic corrections on the CV formula was studied in \cite{sergei},\cite{setare}. One can in a similar manner find out the corrections that the CV formula acquires in light of the higher order corrections that have been derived in these papers.
\paragraph*{}
The holographic principle can be tested by examining the higher order corrections to black hole entropy. The general entropy corrections calculated by the Cardy formula were compared with the corrections to the black hole entropy from canonical considerations in \cite{muin} to test the holographic principle. The correction to the black hole entropy derived in \cite{muin} differ for both the BTZ blackhole and the AdS blackhole. The reason for the same was pointed out earlier. 

\vs{.4cm}
\no
\ul{\textbf{Acknowledgements:}}

\vs{.2cm}
\no

The author thanks Prof Saurya Das for introducing him to this problem and for the discussions he had with him during this work. Also he would like to thank Prof R. K. Bhaduri for his comments and Prof Mark Walton for interesting discussions. The author would also like to thank the referees for pointing out places in the paper that needed explanations. This work was supported by grants from Natural Sciences and Engineering Research, Canada (NSERC) and the University of Lethbridge for which the author shall remain obliged.

\end{document}